\documentclass[conference]{IEEEtran}
\IEEEoverridecommandlockouts
\usepackage{cite}
\usepackage{amsmath,amssymb,amsfonts}
\usepackage{algorithmic}
\usepackage{graphicx}
\usepackage{textcomp}
\usepackage{xcolor}
\def\BibTeX{{\rm B\kern-.05em{\sc i\kern-.025em b}\kern-.08em
    T\kern-.1667em\lower.7ex\hbox{E}\kern-.125emX}}

\usepackage{booktabs} 
\usepackage{soul}
\graphicspath{{figs/}}
\usepackage{braket}
\usepackage{etoolbox}
\makeatletter
\patchcmd{\@maketitle}
  {\addvspace{0.5\baselineskip}\egroup}
  {\addvspace{-0.5\baselineskip}\egroup}
  {}
  {}

\begin{document}

\title{\huge{Invited: Drug Discovery Approaches using \\Quantum Machine Learning}\vspace{-1em}}

\makeatletter
\newcommand{\linebreakand}{%
  \end{@IEEEauthorhalign}
  \hfill\mbox{}\par
  \mbox{}\hfill\begin{@IEEEauthorhalign}
}
\makeatother

%\author{

%\thanks{$^\dagger$These authors contributed equally to this work.}

%\IEEEauthorblockN{Junde~Li$^\dagger$}
%\IEEEauthorblockA{\textit{Department of CSE}\\
%\textit{Penn State University}\\
%jul1512@psu.edu}
%\and
%\IEEEauthorblockN{Mahabubul~Alam$^\dagger$}
%\IEEEauthorblockA{\textit{Department of EE}\\
%\textit{Penn State University}\\
%mxa890@psu.edu}
%\and
%\IEEEauthorblockN{Congzhou~M~Sha}
%\IEEEauthorblockA{\textit{Department of ESM}\\
%\textit{Penn State University}\\
%cms6712@psu.edu}

%\and
%\IEEEauthorblockN{Jian~Wang}
%\IEEEauthorblockA{\textit{College of Medicine}\\
%\textit{Penn State University}\\
%jwang10@psu.edu}

%\linebreakand
%\IEEEauthorblockN{Nikolay~V.~Dokholyan}
%\IEEEauthorblockA{\textit{College of Medicine}\\
%\textit{Penn State University}\\
%dokh@psu.edu}
%\and
%\IEEEauthorblockN{Swaroop~Ghosh}
%\IEEEauthorblockA{\textit{School of EECS}\\
%\textit{Penn State University}\\
%szg212@psu.edu}
%}

\author{

\thanks{$^\dagger$These authors contributed equally to this work.}

\IEEEauthorblockN{Junde~Li$^{1\dagger}$, Mahabubul~Alam$^{2\dagger}$, Congzhou~M~Sha$^{3,4}$, Jian~Wang$^4$, Nikolay~V.~Dokholyan $^{4,5,6}$, Swaroop~Ghosh$^{1,2}$}

\IEEEauthorblockA{
\textit{$^1$Department of Computer Science and Engineering, Penn State University, University Park}\\
\textit{$^2$Department of Electrical Engineering, Penn State University, University Park}\\
\textit{$^3$Department of Engineering Science and Mechanics, Penn State University, University Park}\\
\textit{$^4$Department of Pharmacology, Penn State College of Medicine, Hershey}\\
\textit{$^5$Department of Biochemistry \& Molecular Biology, Penn State College of Medicine, Hershey}\\
\textit{$^6$Departments of Chemistry, and Biomedical Engineering, Penn State University, University Park}\\
\textit{\{jul1512, mxa890, cms6712, jwang10, nxd338, szg212\}@psu.edu}
}
}

\maketitle

\begin{abstract}
Traditional drug discovery pipeline takes several years and cost billions of dollars. 
%to find a single drug molecule. 
Deep generative and discriminative models are widely adopted to assist in drug development. Classical machines cannot efficiently produce atypical patterns of quantum computers which might improve the training quality of learning tasks. We propose a suite of quantum machine learning techniques e.g., generative adversarial network (GAN), convolutional neural network (CNN) and variational auto-encoder (VAE) to generate small drug molecules, classify binding pockets in proteins, and generate large drug molecules, respectively. 
%The results show that quantum GAN is able to reduce up to 98.03\% of parameters with non-undermined representative power, and that quantum CNN achieves 55\% lower cross-entropy loss, and that quantum VAE yields comparable ligand loss with classical counterpart.

\end{abstract}

\section{Introduction}
Modern pharmaceutical research uses automated high-throughput screening technologies to discover new biologically target-binding compounds, but the development of a new drug is still a long and expensive process. 
%Although resources spent on drug discovery have increased exponentially, the number of new drugs approved per year remains constant \cite{convertino2016pharmacological}. 
Computational molecular docking offers an efficient and inexpensive way to identify target-binding compounds and to estimate the binding affinity between compounds and targets. The success rate of virtual drug screening is dictated mainly by 1) the docking accuracy and 2) the comprehensiveness of the compound library used for screening. 
% The comprehensiveness of the compound library determines whether high binding affinity compounds are included in the library, and the docking accuracy determines whether the high binding affinity compounds can be successfully identified by the docking software. 
The docking accuracy of a docking software is decided by its ability to sample compound and target conformations
%\cite{ding2013incorporating,fan2021gpu},
\cite{fan2021gpu},
as well as the accuracy of its scoring method %\cite{yin2008medusascore,jiang2020guiding}.
\cite{jiang2020guiding}.
Significant strides have been made to enhance the sampling and scoring procedures 
%in docking software by developing new scoring and sampling algorithms 
\cite{dagliyan2011structural} and utilizing massive protein–ligand complex structures 
% determined by X-ray, nuclear magnetic resonance spectroscopy, and cryogenic electron microscopy 
to train the scoring function.
%Over the last two decades, 
Numerous docking methods (see Fig. \ref{flow}(a) for docking engine mechanism) have been proposed and evaluated, such as Glide \cite{friesner2004glide}, 
% GOLD \cite{jones1995molecular}, 
MedusaDock \cite{ding2010rapid, wang2019medusadock}, AutoDock Vina \cite{trott2010autodock}. %DOCK \cite{allen2015dock}.
% HADDOCK \cite{dominguez2003haddock}, ICM \cite{totrov1997flexible}, ProDock \cite{trosset1999prodock}, RosettaLigand \cite{meiler2006rosettaligand}, 
% and SwissDock \cite{grosdidier2011swissdock}. Most docking software only support rigid docking: the structural conformation of the receptor remains unchanged in docking; few tools support flexible docking, in which the structural conformation of the receptor is sampled to best fit the ligand. 
%While software such as AutoDock Vina and RosettaLigand support flexible docking, only the conformation of receptor side chains are sampled, and the backbone conformation remain fixed since backbone sampling significantly increases the sampling space, a significant computational hurdle. However, MedusaDock allows for fast receptor backbone sampling.
% In the CSAR 2011 (Community Structure–Activity Resource) Docking Benchmark (www.csardock.org), MedusaDock, equipped with a well-studied scoring function (MedusaScore \cite{yin2008medusascore}) and a flexible side chain and backbone docking protocol, successfully identifies the near-native poses for 28 out of 35 ligands, the highest success rate compared to other methods in near-native pose prediction ($<2.5$ Å RMSD) \cite{ding2013incorporating}.

Quantum computing can offer unique advantages over classical computing in many fields, such as chemistry simulation, machine learning, and optimization. Quantum GAN is one of the main applications of near-term quantum computers due to its strong expressive power in learning data distributions even with much less parameters compared to classical GANs. However, quantum neural network is still at its nascent stage due to qubit constraints on noisy quantum computers. Considering the specific task of drug discovery, we explore potential quantum advantages for both generative and predictive models due to the following reasons: 1) Gate parameter exploration in Hilbert space is different from neural network parameter exploration. Exponentially growing Hilbert space offers chance to explore certain chemical regions that are not accessible to classical GANs. 2) Given a chemical region abundant of molecules, the inherent probabilistic nature of quantum systems helps generate more diverse and novel (though also less valid) molecules surrounding that region.

This paper presents three new Quantum Machine Learning (QML) techniques for drug discovery namely, a hybrid  (i) quantum GAN to learn the patterns in molecular dataset and generate small drug-like molecules, (ii) quantum classifier for protein pocket classification, (iii) quantum VAE to generate a probabilistic cloud of molecules to screen a molecular dataset. %\textbf{Paper organization:} 
In the remaining paper, we cover the background on drug discovery, quantum computing, and QML in Section \ref{back}, discuss three QML applications in drug discovery and development in Section \ref{appr}, and draw conclusion in Section \ref{con}.

\begin{figure*}
\centering
\includegraphics[width=17.5cm]{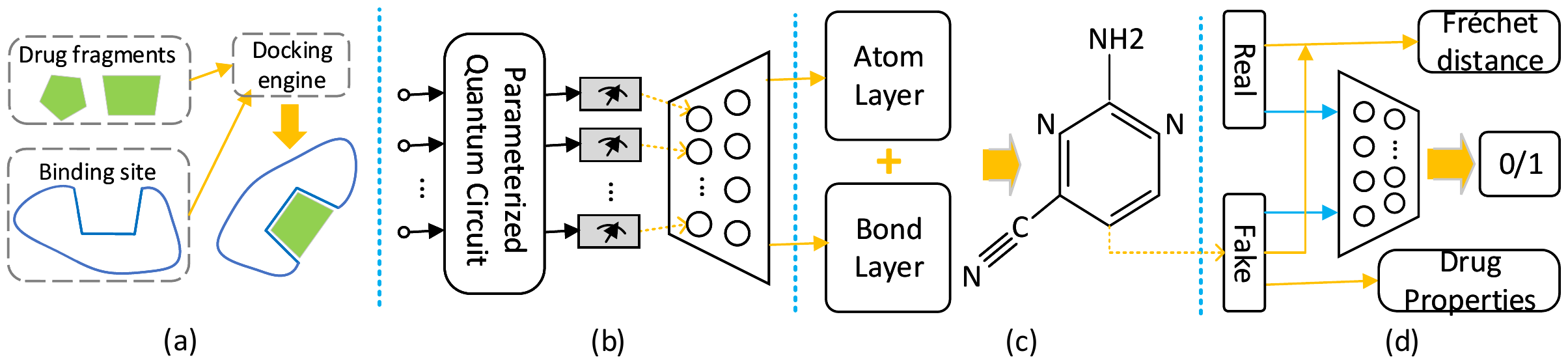}
\vspace{-3mm}
\caption{(a) Only generated molecules that have high affinity towards the receptor binding sites are considered as valid; (b) quantum stage and classical stage separated by blue dotted line; (c) application of atom layer and bond layer for generating synthetic molecular graphs; (d) a batch of real molecules from training dataset (QM9 in this case) and a batch of synthetic molecules generated from (c) are fed into classical discriminator for real/synthetic prediction and FD score calculation, and drug properties for synthetic molecules are evaluated using RDKit \cite{rdkit, jli}. %The prediction losses from discriminator are back-forwarded to two neural networks as well as quantum circuit for updating all parameters simultaneously in each training epoch.
}
\vspace{-5mm}
\label{flow}
\end{figure*}

\section{Background} \label{back}
\textbf{Drug discovery problem: }
Historically, drugs have been discovered incidentally, such as penicillin. % or the discovery of aspirin (acetylsalicylic acid) which was commercialized by Bayer around 1900. 
Nowadays, chemical compounds are selected from databases for extensive preclinical studies based on druglikeness and synthesizability as measured by heuristics. %such as Lipinski's rule of five \cite{ro5}. 
These compounds are screened for biological activity with a variety of protein assays. %, characterizing binding affinity to a given protein and changes to protein enzymatic activity. 
Naturally occurring bioactive compounds may be identified from plant and animal samples from diverse ecosystems.
% , such as paclitaxel from the Pacific yew tree, which targets $\beta-$tubulin in rapidly dividing cells (i.e., cancer cells) \cite{paclitaxel}. 
Medicinal chemists may further optimize selected compounds to maximize binding affinities and synthesizability.
Compounds which show promise in the preclinical setting are then screened in cell lines and animals before making it to human trials. %\cite{drugdev}. 
%Many compounds fail to reach human trials due to safety concerns in these earlier experiments; and even if the compound is successful in animal trials, it may not have the same efficacy and safety in human trials. This high failure rate makes drug discovery an expensive endeavor, which requires a large investment of resources by pharmaceutical companies to successfully bring a drug to market.% Often, it is more cost effective to modify existing drugs than to find novel bioactive compounds, which is a technique commonly used to modify the activity of certain families of drugs. For example, there are five generations of cephalosporins, penicillin-like derivatives, which have varying activities against specific families of bacteria. All five generations contain the characteristic $\beta-$lactam ring which is responsible for the mechanism of action of penicillins, cephalosporins, and carbapenems \cite{betalactam}. With the rise of multidrug-resistant (and in some cases, pandrug-resistant) organisms in the hospital setting, compounds with novel mechanisms of action have become highly desirable \cite{isph}.

\textbf{Classical approaches to drug discovery:}
Compounds are typically experimentally screened for activity against potential receptors \textit{in vitro} by quantitatively measuring the binding of the compound to the receptor. 
%This approach can be done even in the absence of high resolution imaging of the 3D structure, such as in fluorescence resonance energy transfer. %or by measuring enzyme activity in the case that the receptor facilitates a chemical reaction. 
% However, a deeper understanding of the binding to a receptor are usually obtained through X-ray crystallography, NMR imaging, or cryogenic electron microscopy that are often time-consuming and can incur a significant cost (to screen even one receptor against a collection of compounds). 
Computational approaches determine characteristics of receptors and compounds which facilitate binding. The most accurate predictions obtainable come from density functional theory, however these calculations are limited to small molecules and receptor fragments.% and are computationally infeasible for entire receptors.
%\par Molecular dynamics (MD) simulations have long been used to investigate the optimal binding poses of compounds to receptors. In MD, a classical force field is used to approximate the interactions among atoms. Subsequently, a Markov chain Monte Carlo approach may be used to sample poses of both the compound and receptor and predict binding affinities.
\par As an approximation to the quantum mechanical properties of compounds and receptors, a variety of chemical fingerprint algorithms have been developed. %, which characterize and compress information the topology and electron distribution of a molecule into a dense representation. 
Given a molecule and the receptor it binds, fingerprint similarity can be used to screen a library of molecules 
% (e.g., ZINC15 \cite{zinc15}) 
of other compounds which may be active against the receptor.
% These techniques are often used in virtual high-throughput screening (vHTS), which narrow down the compound library for further experimental screening \cite{methods2014}.
Recently, machine learning has become a promising candidate for virtual high-throughput screening pipelines. The MolGAN model uses a generative adversarial network (GAN) to learn and generate drug-like molecules based on the QM9 dataset \cite{de2018molgan, qm9}. To improve the quality of generated compounds and stabilize the GAN training, a reward network based on chemical validity (RDKit\cite{rdkit}) is used. 
% The final trained MolGAN network is composed of fully connected layers with $\tanh$ activation functions. and outputs a sparse representation of an $N$-atom molecule as an $N\times N$ adjacency matrix of bonds between atoms as well as an atom type matrix which predicts each atom's type. %MolGAN is able to generate 2-3\% unique molecules which are not found in the training set, however the model undergoes mode collapse with increased training.

\begin{figure*}
\centering
\includegraphics[width=17cm]{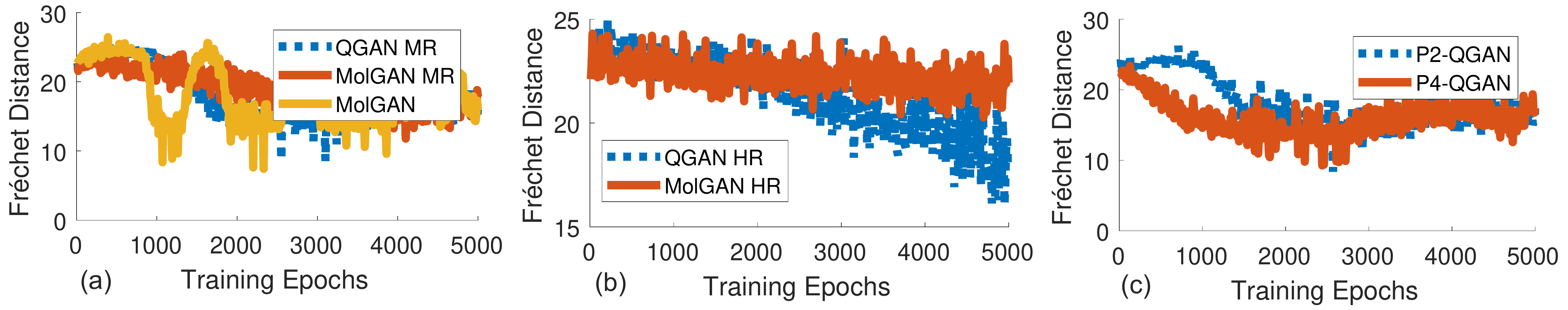}
\vspace{-4mm}
\caption{Training comparison among GAN flavors: (a) Fréchet distances for MolGAN, moderately reduced (14.93\%) MolGAN and QGAN-HG; (b) Fréchet distances for highly reduced (1.97\%) MolGAN and QGAN-HG; (c) learning curves for patched QGAN-HG with two sub-circuits and four sub-circuits \cite{jli}.}
\vspace{-5mm}
\label{training}
\end{figure*}

\textbf{Quantum computing and QML:}
Quantum computing 
%is an orthogonal computing paradigm to classical computation that 
uses quantum phenomena such as, superposition and entanglement to perform computation. Qubits are the building blocks of quantum computers which is analogous to classical bits however, a qubit can be in a superposition state i.e., a combination of 0 and 1 at the same time. Quantum gates (e.g., 1-qubit Pauli-X ($\sigma_x$) gate, 2-qubit CNOT gate, etc.) modulate the state of qubits and thus, perform computation.
% Quantum computers can solve certain computational problems, such as integer factorization substantially faster than classical computers.
% \textbf{Quantum Machine Learning (QML):}
%Machine learning tasks sometimes are hard to learn because of high-dimensional and large-scale data set. Quantum neural networks are essentially are quantum circuits comprising of a series of parameter dependent unitary transformations. Quantum neural network architecture is dependent on qubits, gates, circuit layer, and quantum system configurations. Quantum circuit can either work alone for machine learning tasks or get coupled with classical neural networks considering quantum resource constraints.
QML involves parameter optimization of parameterized quantum circuits to obtain a desired input-output relationship. 
%The inputs/outputs of a QML model can be either quantum or classical. 
% Numerous QML models are already in place, e.g., quantum neural networks \cite{ezhov2000quantum}, quantum generative adversarial network \cite{dallaire2018quantum}, quantum variational autoencoder \cite{khoshaman2018quantum}, quantum convolutional neural networks \cite{cong2019quantum}, etc.
Although none of the existing QML models have a provable performance guarantee over the classical models, several works claim that QML models have high expressibility \cite{du2020expressive}. %It may eventually lead to the superior quantum solutions than the classical ones.

\section{QML for drug discovery} \label{appr}

%\hl{will put some introduction: Mahabubul}

\begin{figure*} 
 \begin{center}
    \includegraphics[width=0.99\textwidth]{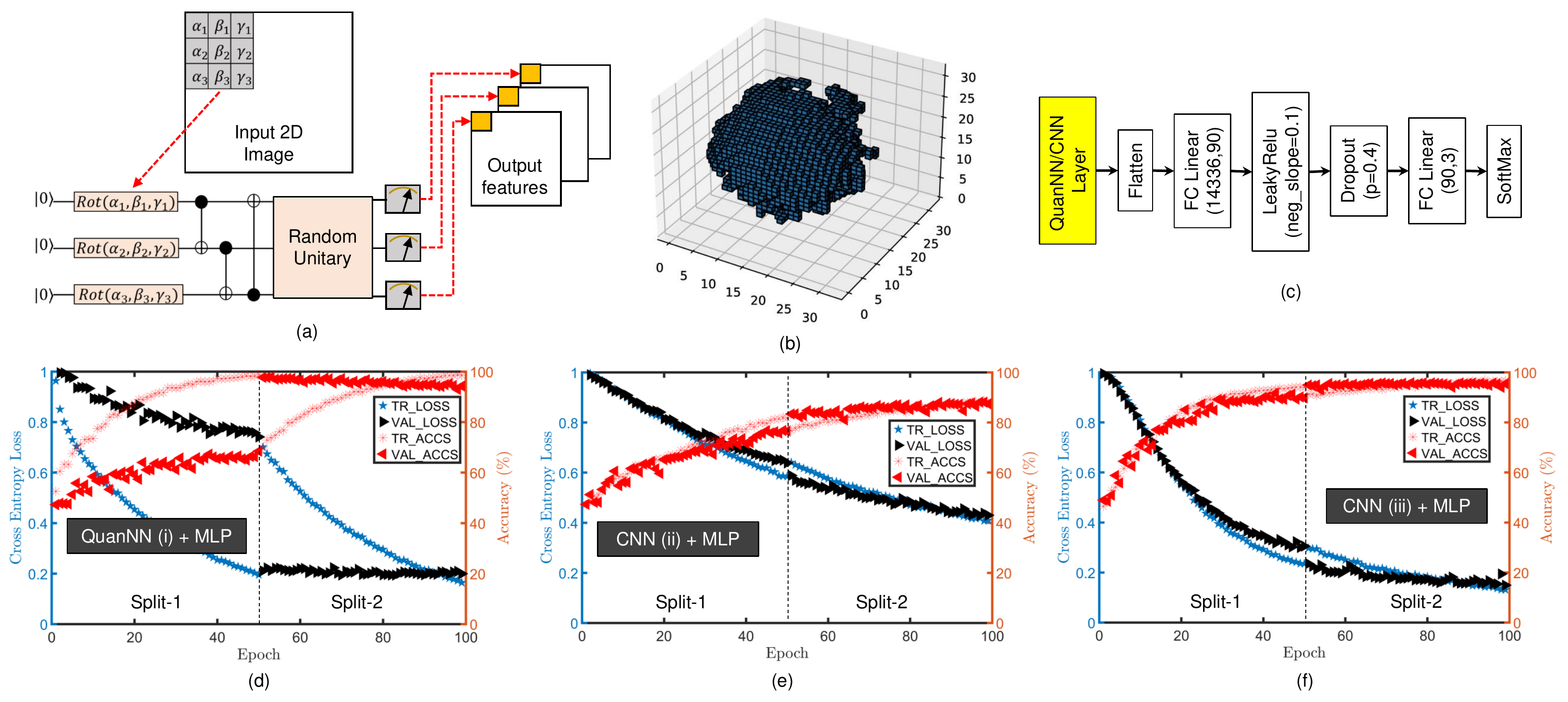}
 \end{center}
 \vspace{-6mm}
\caption{(a) A toy Quanvolution operation on a 2D image segment to generate output feature maps using random quantum circuit, (b) a single channel of a sample protein pocket in the TOUGH-C1 dataset, (c) deep learning pipeline used in the case-study, (d) performance of QuanvNN (no trainable parameters) + MLP, (e) CNN (no trainable parameters) + MLP, and (f) CNN (with trainable parameters) + MLP.} \label{fig:deepdrug}
\vspace{-5mm}
\end{figure*}

\subsection{Approach-1: QGAN-HG}

Generative learning with graph-structured molecules is invariant to the orderings of atoms \cite{de2018molgan} and automates the navigation to a chemical region abundant in desired molecules.
% Drug molecules can be represented as graphs where the nodes and edges correspond to atoms and bonds, respectively. Given the task complexity of learning molecule distribution, full quantum GAN can hardly encode all training data quantum mechanically. Take the small molecule dataset QM9 \cite{qm9} for example. The total number of qubits required for reconstructing synthetic molecules is $\binom{9}{2}\log5 + 9\log5 > 90$ where 5 is the number of bond types and atom types contained in QM9. No commercially available quantum computers support over 90 qubits at present. 
The proposed qubit-efficient quantum GAN with hybrid generator and classical discriminator (QGAN-HG) \cite{jli} efficiently learns molecule distributions based on classical MolGAN (Fig. \ref{flow}). We also examine the patched circuit idea \cite{pan} by comparing to the original single large generator circuit implementation using metric of Fréchet distance and drug property scores. %Fig. \ref{flow} shows the overall workflow of the qubit-efficient hybrid quantum GAN model for drug discovery. 
% Quantum GAN has a few flavors of the generator and discriminator implementations depending on their execution environments, either on quantum computers, classical machines or quantum simulators. However, only QGAN-HG is applicable to learning complicated tasks like drug discovery due to qubit resource constraint.
Quantum GAN with hybrid generator (QGAN-HG) is composed of a parameterized circuit to get a feature vector of qubit size dimension, and a classical deep neural network to output an atom vector and a bond matrix for the graph representation of drug molecules. Another patched quantum GAN with hybrid generator (P-QGAN-HG) is considered as the variation of QGAN-HG where the quantum circuit is formed by concatenating few quantum sub-circuits. The variational quantum circuit consists of 3 stages, namely initialization, entanglement and measurement stages. 
% The initialization stage takes value randomly sampled from $[-\pi, \pi]$ to substitute the random Gaussian noise input for classical GANs. Let us denote the parameterized layers in entanglement stage repeated for $L$ times as unitary matrix $U(\vtheta)$. The final quantum state is of the form $\ket{\Psi(z)} = U(\vtheta)\ket{z}$. 
Classical stage of hybrid generator is a standard neural network with input layer receiving the feature vector of expectation values. Note that, 85.07\% and 98.03\% of generator parameters are respectively cropped from classical GAN \cite{de2018molgan} to demonstrate the strong expressive power of quantum circuits. %To measure the effects of circuit layer and qubit count, 
We also implement two enhanced QGAN-HG variants with $L=2$ and $N=10$.

We conduct the experiments with QM9 \cite{qm9} dataset.
% which contains 133,885 molecules with up to 9 heavy atoms of types of carbon, nitrogen, oxygen, and fluorine. 
%Learning results of the proposed GANs are evaluated with Fréchet distance metric which measures the similarity between real and generated molecule distributions. 
Generated molecule distribution is approximately created by generating a batch of molecules, and real one is approximately formed by randomly sampling the same number of molecules from QM9. Generated drug quality is evaluated using metrics such as druglikeness, solubility and synthesizability using RDKit \cite{rdkit}. %We pivot on the classical MolGAN \cite{de2018molgan} to implement our QGAN-HG and P-QGAN-HG algorithms. 
% Our QGAN variations are trained with a mini-batch of 128 molecules using Adam optimizer on a single RTX 2080 Ti GPU for classical part and PennyLane platform \cite{pennylane}. 
The learning rate is set to 0.0001 %for both generator and discriminator 
and starts decaying uniformly at a factor of $1/2000$ after 3000 epochs. Total training epoch is set with 5000, and early stopping based on Fréchet distance is applied under model collapse.

Fig. \ref{training}(a-b) compares the performance between MolGAN and QGAN-HG for moderately and highly reduced architectures. All mechanisms can reach a reasonably good training point within 5000 epochs, however, moderately reduced MolGAN takes around 4000 iterations while original MolGAN and QGAN-HG take $\approx$2500. MolGAN with highly reduced architecture can hardly be learned though a slight downward trend is observed. Quantum circuit involves only 15 gate parameters illustrating the strong expressive power of variational quantum circuits. We examine two patched QGAN-HG variations, i.e., P2-QGAN with two sub-circuits (each has 4 qubits and 7 gate parameters) and P4-QGAN with four sub-circuits (each has 2 qubits and 3 parameters). Surprisingly, the learning quality of the patched QGANs (with even less gate parameters) are comparable to QGAN with an integral circuit (Fig. \ref{training}(a, c)). 
% Further, the simulation time (see Fig. \ref{training}(f)) for patched quantum circuits are significantly reduced because of smaller qubit count and early convergence.

\subsection{Approach-2: Image based search}

%Machine learning (particularly deep learning) has emerged as a crucial tool in modern drug discovery and development flows. 
% Successful demonstrations of classical ML models in protein fold predictions \cite{hou2018deepsf} (\emph{DeepSF}), protein-binding site predictions \cite{jimenez2017deepsite} (\emph{DeepSite}), ligand chemical property predictions \cite{skalic2019ligvoxel} (\emph{LigVoxel}) and protein-ligand affinity predictions \cite{jimenez2018k} (\emph{KDEEP}) exist. %may move the above paragraph later
Most of the deep learning solutions for drug discovery use 3D grid representation of the molecular structures. Each grid-point is commonly referred to as \emph{voxels}.
% \footnote{The steps involved in voxelization of any protein structure available in the protein data bank \cite{rose2015rcsb} is beyond the scope of this paper. Interested readers can look into the following articles: \cite{skalic2019ligvoxel, pu2019deepdrug3d}.}. 
For example, voxel representation of ligand-binding protein pockets are used as inputs to a convolutional deep neural network (CNN) in \cite{pu2019deepdrug3d} to classify the pockets in one of 3 groups: \emph{nucleotide-binding} or \emph{heme-binding} or \emph{others}. 
%The field of quantum machine learning (QML) is advancing at a tremendous pace. Numerous QML models are already in place, e.g., quantum neural networks \cite{ezhov2000quantum}, quantum generative adversarial network \cite{dallaire2018quantum}, quantum variational autoencoder \cite{khoshaman2018quantum}, quantum convolutional neural networks \cite{cong2019quantum}, etc.
%Although none of the existing QML models have a provable performance guarantee over the classical models, several works claim that QML models have high expressibility \cite{du2020expressive, sim2019expressibility}. It may eventually lead to the development of quantum solutions that outperform all the classical ones. Observing these trends, we believe that QML models will soon be part of the conventional deep learning pipelines.
%The field of quantum machine learning (QML) is advancing at a tremendous pace. Observing these trends, we believe that QML models will soon be part of the deep learning pipelines. 
In this section, %we present a case-study on a QML model performance in a deep learning pipeline for a drug discovery and development related task compared to the fully classical approach. 
we pick the pocket classification problem in \cite{pu2019deepdrug3d} as our test-case and the \emph{quanvolutional neural networks} presented in \cite{henderson2020quanvolutional} to gauge the efficacy of QML models in drug development pipelines.
% The details are discussed below.

{\textbf{Quanvolutional Neural Networks:}} A quantum analogous of the conventional CNN for image recognition tasks has been proposed \cite{henderson2020quanvolutional} (referred as Quanvolutional Neural Networks (QuanNN)) where filters are replaced by quantum circuits. Each filter takes a segment of the input image and encode the data as a quantum state with suitable encoding scheme (e.g., angle or amplitude encoding). Later, a random (/structured) quantum circuit transforms the state and the output is measured repeatedly to generate a feature map. % using suitable decoding scheme (e.g., Pauli-Z expectation value of the qubits). 
Similar to CNN, the filter moves across the entire image in finite steps to generate new feature maps for the entire image (Fig. \ref{fig:deepdrug}(a)). %The transformed data is used as inputs to the downstream networks. %A toy quanvolution step on a 2D image segment is shown in Figure \ref{fig:deepdrug}(a).

{\textbf{Dataset:}} The TOUGH-C1 dataset \cite{pu2019deepdrug3d} used for this work contains 4095 samples. Each sample is represented by a 14x32x32x32 tensor (14 channels). The dataset contains 1553 nucleotide-binding, 596 heme-binding, and 1946 other type of protein pockets. A single channel of a protein pocket in the dataset is shown in Fig. \ref{fig:deepdrug}(b).

{\textbf{Deep Learning Pipeline:}} We have used 3 deep learning pipelines with QuanNN, CNN, and multi-layer perceptrons (MLP): (i) single-layer QuanNN (no trainable parameters) + MLP, (ii) single-layer CNN (no trainable parameters) + MLP, and (iii) single-layer CNN + MLP (Fig. \ref{fig:deepdrug}(c)). %The MLP consists of two fully connected layers. The hidden layer (90 neurons) is followed by non-linear 'LeakyRelu' activation (with a negative-slope of 0.1). A 'dropout' layer is used after 'LeakyRelu' for regularization (p=0.4). The output layer (3 neurons) is followed by 'SoftMax' activation functions. 
Cross entropy is used as the loss function for all these networks.
The QuanNN and CNN layers both produce 14336 dimensional representation of the input 3D pocket grids. For the CNN, we use 28 output channels with kernel-size = 4 and, and stride = 4 which produces 8x8x8x28 (14336) dimensional output features. For the QuanNN, we use a 4-qubit quantum circuit as the filter and take the Pauli-Z expectation values of the qubits (4) as output features. 
The quantum circuit consists of 'StronglyEntanglingLayers' from Pennylane \cite{pennylane} that encodes the input classical data as rotation angles at different qubits followed by all-qubit entanglement with CNOT operations. A random 4-qubit quantum circuit is placed in front of the encoding block for random transformation of the quantum state using 'RandomLayers' class in the Pennylane framework. The quantum circuit takes a 4x4x4 dimensional tensor block from 2 channels at a time (2x4x4x4 dimensional input). For 14x32x32x32 dimensional 3D grids (14 channels), it produces 7x8x8x8x4 (14336) output features. %when the filter is moved across the grid without overlap. %For QuanNN, we normalize the input data with $\mu = 0$, and $\sigma = 1$.

{\textbf{Metrics:}} We model and train all three networks using PyTorch and Pennylane libraries under identical configurations. We perform stratified 2-fold cross-validation training of the models for 50 epochs/split with a batch-size of 32 and learning-rate of $e^{-5}$. We use the cross-entropy loss and classification accuracy as the metrics for performance. 

{\textbf{Comparison:}} In our simulations, (i) outperformed (ii) by a significant margin (Fig. \ref{fig:deepdrug}(d)\&(e)). At the end of the training, (i) achieved $\approx$55\% lower cross-entropy loss, and $\approx$17\% higher accuracy over (ii) for the training set. For the validation set, the loss was $\approx$57\% lower, and accuracy was $\approx$14\% higher. However, compared to (iii) (Fig. \ref{fig:deepdrug}(f)), (i) showed $\approx$11\% higher loss but $\approx$2\% higher accuracy for the training set and $\approx$15\% higher loss and $\approx$1\% lower accuracy for the validation set. Note that, (iii) has trainable parameters in the CNN layer which are learned during the training alongside the trainable parameters in the MLP. These additional CNN parameters result in a better performance. We could have added trainable parameters at the quantum circuit in quanvolutional layer at the cost of prohibitive training time overhead.

The comparison between (i) and (ii) is more rational as they both perform random transformation of the input data - a technique often used in classical ML frameworks \cite{rahimi2007random}. In both cases, the input features are transformed to a different feature space using a random kernel function. %In (ii), the CNN filters with randomly initialized weights act as the kernel function whereas, in (i), the random quantum circuit performs the kernel operations. %Note that the performance benefits observed here matches with the results presented in \cite{henderson2020quanvolutional} which was performed in a much smaller scale on the MNIST dataset with a different ad-hoc encoding-decoding approach.

%\hl{may move to future outlook: Mahabubul}
%Although all the quanvolution operations are highly parallelizable, performing them in a classical machine is extremely memory- and compute-intensive. For instance, in this work, we required 7x8x8x8 or 3584 quanvolution operations (random quantum circuit kernel simulation) per input. A density matrix simulator with float32 precision requires $2^4$x$2^4$x32 or 8KB memory to store the density matrix of a single quanvolution kernel operation. For the 3584 kernel operations, we require 28 MB of data just to store the density matrix. For an entire batch of 32 pockets, this accounts to 896 MB of memory. Additionally, significant memory needs to be allocated to assist in gradient computation during back-propagation. The entire process can easily go beyond the capabilities of classical simulation if we choose bigger kernels with higher number of qubits. Contrarily, all these 4-qubit (or bigger) circuits can be run in parallel in a single-large or cluster of quantum computers quite efficiently (e.g. a 3584x4x32 qubit quantum computer or a quantum cluster could perform all the necessary quanvolutional operations concurrently within a few milliseconds). Therefore, such methods can be more practical in the near future when we have access to sufficient quantum resources.  

\subsection{Approach-3: Probabilistic search}
In this approach, we treat the receptor as an $N\times N\times N$ image, with $8$ channels corresponding to the atom types $\{\emptyset, C, N, O,S,P,H,X\}$, and $X$ is an atom other than those listed. The ligand is treated as an $N\times N$ adjacency matrix, with 6 channels corresponding to the bond types $\{\emptyset, \text{SINGLE},\text{DOUBLE},\text{TRIPLE},\text{AROMATIC}\}$; and a length $N$ atom type vector, with 7 channels corresponding to the atom types $\{\emptyset, C, N, O, S, F, X\}$. Probability distributions for active compounds are generated for given receptor pocket image. Note that $N=32$ sets the maximum number of atoms, making the approach feasible for generating large molecular graphs.

\begin{figure}
\centering
\includegraphics[width=8cm]{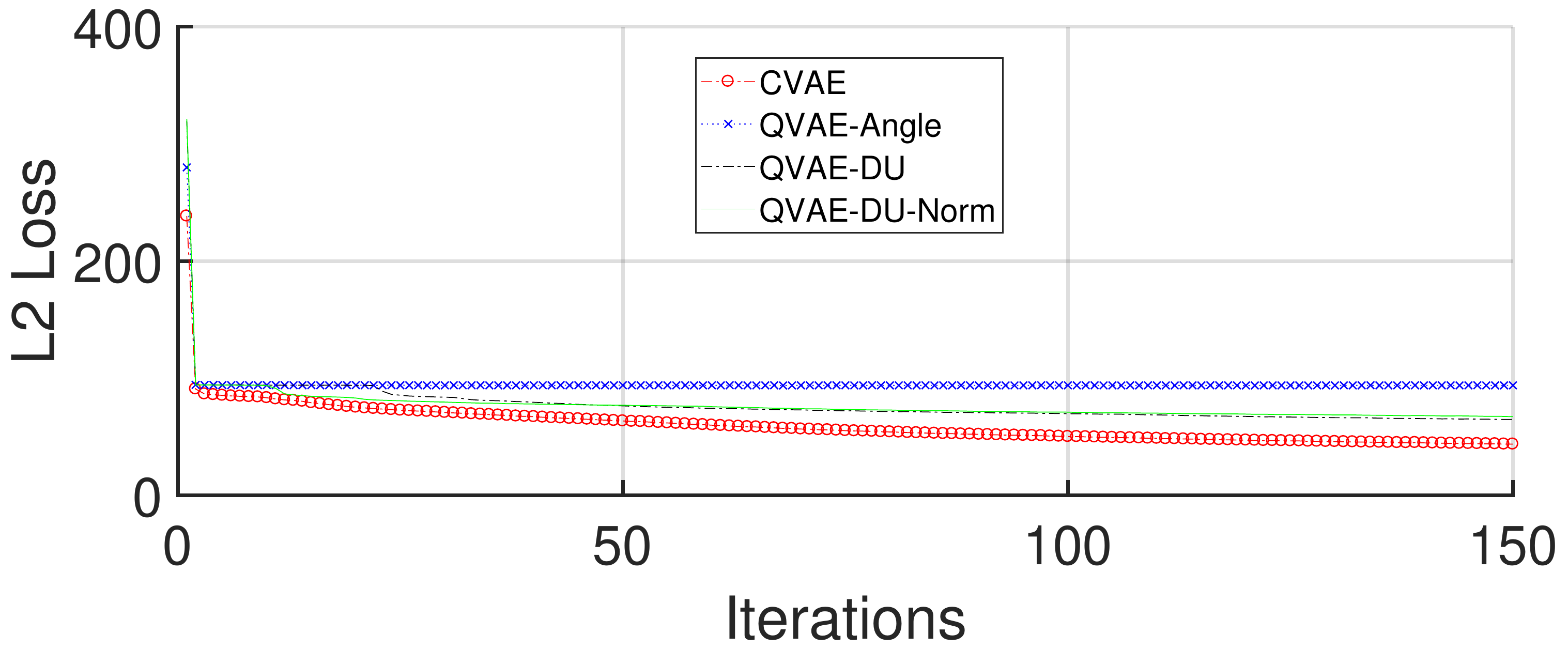}
\vspace{-4mm}
\caption{Learning results for classical ligand VAE, quantum VAE with angle embedding, data uploading, and its normalized version.}
\vspace{-6mm}
\label{qvae}
\end{figure}

\par Two variational autoencoders (VAE), one for the receptor pocket and one for the ligand are constructed for matching pairs of receptor pocket and ligand. The receptor VAE consists of 3D convolutional layers, compressing the original image $R_0$ into a 32-element latent representation $R_c$ and decompressing back into the original receptor pocket image $R_d$. We represent these functions as $E_R(R_0)=R_c$, $D_R(R_c)=R_d$. Similar ligand VAE architecture is applied except for non-convolutional layers. Two VAEs are connected with a two-way matching network, which converts the latent representation of the receptor pocket into the latent representation of the ligand ($M_{R\rightarrow L}: R_c\rightarrow L_c$), and vice versa ($M_{L\rightarrow R}: L_c\rightarrow R_c$). The two directions of the matching network are independent.

% We train our matching network with the following loss:
% $$\mathcal{L}_{\text{match}}=KL(L_0\big\vert (D_L\circ M_{R\rightarrow L}\circ E_R)(R_0))$$
% $$+KL(R_0\big\vert (D_R\circ M_{L\rightarrow R}\circ E_L)(L_0))$$
% Given the following losses:
% $$\mathcal{L}_{\text{R}}=KL(R_0\big\vert (D_R\circ E_R)(R_0))$$
% $$\mathcal{L}_{\text{L}}=KL(L_0\big\vert (D_L\circ E_L)(L_0))$$
% we train our VAE for the receptor with
% \begin{equation}
%     \mathcal{L}_{VAE, R}=\mathcal{L}_{\text{R}}+\mathcal{L}_{\text{match}}
% \end{equation}
% the VAE for the ligand with
% \begin{equation}
%     \mathcal{L}_{VAE, L}=\mathcal{L}_{\text{L}}+\mathcal{L}_{\text{match}}
% \end{equation}

For all layers, we use LeakyReLU activation functions and wrap the weights with spectral normalization to prevent exploding/vanishing gradients. Hybrid VAE is created by inserting quantum layers at certain positions in neural networks. Quantum VAE performance are possibly affected by factors, such as quantum embedding techniques, and types of quantum layers. We adopt two currently available quantum embedding techniques i.e., angle embedding and data re-uploading method \cite{data-reuploading} for measuring different techniques for quantum state preparation. Furthermore, the impact of adding normalization layer immediately after quantum circuit is compared with non-normalized one. Results in Fig. \ref{qvae} show that QVAE with angle embedding performs poorly, while the classical VAE better in terms of learning~speed. Normalized quantum layer helps at the beginning but the effect diminishes as training runs longer. Since no benefit is observed for quantum VAE, we assume the quantum embedding stage restricts the quantum superiority.

\section{Conclusion and outlook} \label{con}
%At current stage, quantum neural networks show learning benefits only under specific conditions. 
We propose drug discovery techniques using QML. We show quantum superiority over classical GAN in terms of architecture complexity, and over classical CNN under the non-parameterized setting, but no superiority is observed over classical VAE. The challenges of QML lie in designing qubit-efficient differentiable quantum embedding technique and developing quantum circuits catered for specific learning tasks.

\textbf{Acknowledgements:} The work is supported in parts by NSF (CNS-1722557, CCF-1718474, OIA-2040667, DGE-1723687 and DGE-1821766) and seed grants from Penn State ICDS and Huck Institute of the Life Sciences. NVD acknowledges support from the National Institutes for Health (1R35 GM134864) and the Passan Foundation. We also thank Mehrdad Mahdavi and Rasit Topaloglu for helpful discussions.

\bibliographystyle{IEEEtran}
\bibliography{biblio}

\end{document}